\documentstyle[12pt]{article}
\def \be{\begin{equation}}
\def \ee{\end{equation}}
\def \lb{\label}
\def \ba{\begin{array}{l}}
\def \ea{\end{array}}

\def \t{\tau}
\def \a{\alpha}

\def \d{\delta}
\def \D{\Delta}
\def \e{\epsilon}
\def \f{\phi}
\def \g{\gamma}

\def \lm{\lambda}
\def \n{\nabla}
\def \p{\varphi}
\def \tl{\tilde}
\def \ol{\overline}

\def \l{\left}
\def \la{\langle}
\def \ra{\rangle}

\def \fr{\frac}
\def \R{R_{c}}
\def \T{T_{c}}

\def \I{\int d^{D}x}

\def \2{\frac{1}{2}}
\def \4{\frac{1}{4}}

\newcommand{\sectio}[1]{\section{#1}\setcounter{equation}{0}}

\begin{document}

\begin{center}

{\Large \bf Replica Symmetry Breaking and the Renormalization Group Theory
of the Weakly Disordered Ferromagnet}

\vskip .4in

Viktor Dotsenko$^{1,2}$ and D.E.Feldman$^{1}$

\vskip .2in

1. Landau Institute for Theoretical Physics,\\
Russian Academy of Sciences, \\
Kosygina 2, 117940 Moscow, Russia\\

\vskip .2in

2. Laboratoire de Physique Theorique,\\
Ecole Normale Superieure,\\
24 rue Lhomond, 75231 Paris, France\\

\end{center}

\vskip 1in

\begin{abstract}
We study the critical properties of the weakly disordered $p$-component
ferromagnet in terms of the renormalization group (RG) theory generalized
to take into account the replica symmetry breaking (RSB) effects coming
from the multiple local minima solutions of the mean-field equations.
It is shown that for $p < 4$ the traditional RG flows at dimensions
$D=4-\e$, which are usually considered as describing the disorder-induced
universal critical behavior, are unstable with respect to the RSB potentials
as found in spin glasses. It is demonstrated that for a general type of the
Parisi RSB structures there exists no stable fixed points, and the RG flows
lead to the {\it strong coupling regime} at the finite scale
$R_{*} \sim \exp(1/u)$, where $u$ is the small parameter describing the
disorder. The physical concequences of the obtained RG solutions are
discussed. In particular, we argue, that discovered RSB strong coupling
phenomena indicate on the onset of a new spin glass type critical behaviour
in the temperature interval $\t < \t_{*} \sim \exp(-\fr{1}{u})$ near $\T$.
Possible relevance of the considered RSB effects for the Griffith phase
is also discussed.
\end{abstract}
\newpage

\sectio{Introduction}

In this paper we study the effects produced by weak quenched
disorder on the critical phenomena
in the ferromagnetic spin systems near the phase transition point.
In the most general terms the traditional
point of view on this problem could be summarized as follows.

According to the usual scaling theory near the critical temperature $\T$
the only relevant scale that remains in the
system is the correlation length $\R$ which scales as
$\sim \tau^{-\nu}$, where $\t \equiv (T-\T)/\T << 1$ is the reduced
temperature parameter and $\nu$ is the correlation length critical
exponent.

If the disorder is weak (e.g. the concentration of impurities
is small), its effect on the critical behavior in the vicinity of
the phase transition point $\T$ remains negligible so long as the
correlation length $\R$ is not too large, i.e. for temperatures
$T$ not too close to $T_{c}$. In this regime the critical behavior
will be essentially the same as in the pure system.

However, in the close vicinity of the critical point, at
$\t \equiv (T-\T)/\T \to 0$, the correlation length $\R$ grows and
becomes larger than the average distance between the impurities,
so that the effective concentration of impurities, measured with
respect to the correlation length, becomes large.
The strength of disorder, described by small parameter $u$, affects
only the width of the temperature region near $\T$ in which the
effective concentration gets large. If $u \R^{D} \gg 1$, where
$D$ is the spatial dimensionality, one has no grounds, in
general, for believing that the effect of impurities will be
small.

A very simple general criterion has been discovered, the so-called
Harris criterion \cite{harris}, which makes it possible to predict
the effect of impurities qualitatively from only the critical
exponents of the pure system.  According to this criterion the
impurities change the critical behavior only if $\alpha$, the
specific heat exponent of the pure system, is greater than zero
(i.e. the specific heat of the pure system is divergent at the
critical point). According to the traditional point of view, when this
criterion is satisfied, the disorder becomes relevant and a new universal
critical behavior, with new critical exponents, is established sufficiently
close to the phase transition point \cite{new,newnew}:

\be
\lb{aaam}
\t < \t_{u} \equiv u^{1/\a}
\ee
This argument identifies $1/\alpha$ as the cross-over exponent associated
with randomness \cite{AA}. In contrast, when $\a < 0$ (the
specific heat is finite), the disorder appears to be irrelevant,
i.e. their presence does not affect the critical behavior.

Near the phase transition
point the $D$-dimensional Ising-like systems can be described in
terms of the scalar field Ginsburg-Landau Hamiltonian with a
double-well potential:

\be
\lb{aaaa}
H = \I \Biggl[ \2 (\n\f(x))^{2} + \2 [\t - \d\t(x)] \f^{2}(x) +
 \frac{1}{4}g \f^{4}(x) \Biggr] \ .
\ee
Here the quenched disorder is described by random fluctuations of
the effective transition temperature $\d\t(x)$ whose probability
distribution is taken to be symmetric and Gaussian:

\be
\label{aaab}
P[\d\t] = p_{0} \exp \Biggl( -\frac{1}{4u}\I (\d\t(x))^{2} \Biggr) \ ,
\ee
where $u \ll 1$ is the small parameter which describes the disorder,
and $p_{0}$ is the normalization constant.  In Eq. (\ref {aaaa})
$\tau \sim (T-T_c)$ and for notational simplicity, we define the sign
of $\delta \tau (x)$ so that positive fluctuations lead to
locally ordered regions.

Now, if one is interested in the critical properties of the
system, one has to integrate over all local field
configurations up to the scale of the correlation length.
This type of calculation is usually performed using a
Renormalization Group (RG) scheme, which self-consistently
takes into account all the fluctuations of the field on
length scales up to $\R$.

To derive the traditional results for the critical properties of this system
discussed above one can use
the usual RG procedure developed for dimensions $D = 4 - \e$, where $\e \ll 1$.
Then one finds that in the presence of the quenched disorder the
pure system fixed point becomes unstable,
and the RG rescaling trajectories are arriving to another (universal)
fixed point $g_{*} \not= 0$; $u_{*} \not= 0$, which yields the new
critical exponents describing the critical properties of the system
with disorder.

However, there exists an important point which missing in the
traditional approach. Consider the ground state properties of the
system described by the Hamiltonian (\ref{aaaa}).
Configurations of the fields $\f(x)$ which correspond to local
minima in $H$ satisfy the saddle-point equation:
\be
\lb{aaac}
-\D\f(x) + (\t - \d\t(x))\f(x) + g\f^{3}(x) = 0 \ .
\ee
Clearly, the solutions of this equations depend on a particular
configuration of the function $\d\t(x)$ being inhomogeneous.
The localized solutions with non-zero value of $\f$ exist in regions
of space where $\tau - \delta \tau (x)$ has negative values.
Moreover, one finds
a {\it macroscopic} number of local minimum solutions of the
saddle-point equation (\ref{aaac}).  Indeed, for a given
realization of the random function $\d\t(x)$ there exists
a macroscopic number of spatial "islands" where
$\tau - \d\t(x)$ is negative (so that the local effective
temperature is below $\T$), and in each of these "islands"
one finds two local minimum configurations of the field:
one which is "up", and another which is "down".
These local minimal energy configurations are separated by
finite energy barriers, whose heights become larger
as the size of the "islands" are increased.

The problem is that the traditional RG approach is
only a perturbative theory in which
one treats the deviations of the field around the ground
state configuration, and it can not take into account
other local minimum configurations which are "beyond barriers".
This problem does not arise in the pure systems, where the
solution of the saddle-point equation is unique. However, in
a situation like that discussed above, when one gets numerous
local minimum configurations separated by finite barriers, the
direct application of the traditional RG scheme may be questioned.

In a systematic approach one would like to integrate in an
RG way over fluctuations around the local minima configurations.
Furthermore, one also has to sum over all these local minima up
to the scale of the correlation length. In view of the fact that
the local minima configurations are defined by the random
quenched function $\d\t(x)$ in an essentially non-local way, the
possibility to implement successfully such a systematic approach
seems rather hopeless.

On the other hand there exists another technique which has been
developed specifically for dealing with systems which exhibit
numerous local minima states.  It is the Parisi Replica Symmetry
Breaking (RSB) scheme which has proved to be crucial in the
mean-field theory of spin-glasses (see e.g. \cite{sg}). Recent
studies show that in certain cases the RSB approach can also be
generalized for situations where one has to deal with
fluctuations as well \cite{manifold},\cite{rsb-Marc},
\cite{Korshunov}.

It can be argued that the summation over multiple local minima configurations
in the present problem could provide additional non-trivial RSB interaction
potentials for the fluctuating fields \cite{dhss}.
Let us consider this point in some more details.

To carry out the appropriate average over quenched disorder we
can use the standard replica approach.
To do this, we need to average the $n$th ($n \rightarrow 0$) power
of the partition function.  This is accomplished by introducing
the replicated partition function, $Z_n \equiv \ol{Z^{n}[\d\t]}$,
where $\ol{(...)}$ denotes the averaging over $\d\t(x)$ with the probability
distribution (\ref{aaab}). Simple integration yields:

\be
\label{bbbd}
\begin{array}{l}
Z_{n} \equiv \ol{Z^{n}(\d\t)} = \int D\f_{a}(x)
\exp \Biggl[ - \I \Biggl(
\2\sum_{a=1}^{n}[\n\f_{a}(x)]^{2}
+ \2\t\sum_{a=1}^{n}\f^{2}_{a}(x) \\
\\
+ \4 \sum_{a,b=1}^{n} g_{ab}
\f^{2}_{a}(x) \f^{2}_{b}(x)] \Biggr) \Biggr] \ ,
\end{array}
\ee
where
\be
\lb{bbbe}
g_{ab} = g\d_{ab} - u \ .
\ee
is the {\it replica symmetric} (RS) interaction parameter. If one
would start the usual RG procedure for the above replica Hamiltonian
(as it is done in the traditional approach), then it would
correspond to the perturbation theory around the homogeneous
ground state $\f = 0$.

However, in the situation when there exist numerous local minima
solutions of the saddle-point equation (\ref{aaac}) one has to be more careful.
Let us denote the local solutions of the eq.(\ref{aaac})
by $\psi^{(i)} (x)$ where $i=1,2,\dots N_0$ labels the "islands"
where $\d\t(x) > \t$.  If the size $L_0$ of an
"island" where $(\d\t(x) - \t) > 0$ is not too small, then the value of
$\psi^{(i)} (x)$ in this "island" should be $\sim \pm \sqrt{(\d\t(x)-\t)/g}$,
where $\delta \tau(x)$ should now be interpreted as the value
of $\delta \tau$ averaged over the region of size $L_0$. Such
"islands" occur at a certain finite density per unit volume.
Thus the value of $N_0$ is macroscopic:
$N_0 = \kappa V$, where $V$ is the volume of the system and
$\kappa$ is a constant.  An approximate global extremal solution
$\Phi (x)$ is constructed as the union of all these local
solutions without regard for interactions between "islands."
Each local solution can occur with either sign, since we are
dealing with the disordered phase:

\be
\lb{PSI}
\Phi_{(\a)} [x;\d\t(x)] = \sum_{i=1}^{\kappa V}
\sigma_i \psi^{(i)} (x) \ ,
\ee
where each $\sigma_i = \pm 1$.
Accordingly, the total number of global solutions must be $2^{\kappa V}$.
We denote these solutions by $\Phi_{(\a)}[x;\d\t(x)]$,
where $\alpha = 1, 2,...,K = 2^{\kappa V}$.
As we mentioned, it seems unlikely that an integration
over fluctuations around $\phi(x)=0$ will include the
contributions from the configurations of $\phi(x)$ which
are near a $\Phi(x)$, since $\Phi(x)$ is "beyond a
barrier," so to speak.  Therefore, it seems appropriate
to include separately the contributions from small fluctuations
about each of the many $\Phi_{(\a)}[x;\d\t]$.
Thus we have to sum over the $K$ global minimum solutions
(non-perturbative degrees of freedom)
$\Phi_{(\a)}[x;\d\t]$ and also
to integrate over "smooth" fluctuations $\p(x)$ around them

\be
\lb{cccd}
\begin{array} {l}
Z[\d\t] = \int D\p(x) \sum_{\a}^{K}
\exp \Biggl(  - H \Biggl[
\Phi_{(\a)}[x;\d\t] + \p(x); \d\t \Biggr] \Biggr) \\
\\
= \int D\p(x) \exp \Biggl( -H[\p;\d\t ] \Biggr)
\times \tl{Z}[\p;\d\t] \ ,
\end{array}
\ee
where

\be
\lb{ccce}
\begin{array} {l}
\tl{Z}[\p;\d\t] = \sum_{\a}^{K}
\exp \Biggl( -H_{\a}  -\I \Biggl[
\frac{3}{2}g  \Phi_{(\a)}^{2} \p^{2}(x) +
g\Phi_{(\a)}\p^{3}(x) \Biggr] \Biggr) \ .
\end{array}
\ee
and $H_{\a}$ is the energy of the $\a$-th solution.

Next we carry out the appropriate average over quenched disorder, and for
the replica partition function, $Z_n$, we get:

\be
\lb{cccf}
Z_{n} = \int D\d\t \int D\p_a \exp \left(
 -\fr{1}{4u}\I [ \d\t(x) ]^{2}
- \sum_{a=1}^{n} H[\varphi_a;\d\t] \right)
\times \tl{Z_{n}}[\p_a;\d\t] \ ,
\ee
where the subscript $a$ is a replica index and

\be
\lb{cccg}
\tl{Z_{n}}[\p_{a};\d\t] =
\sum_{\a_{1}...\a_{n}}^{K}
\exp \Biggl( -\sum_{a}H_{\a_{a}} - \I \sum_{a}
\Biggl[ \frac{3}{2}g \Phi_{(\a_{a})}^{2}(x) \p_{a}^{2}(x) +
g\Phi_{(\a_{a})}(x) \p_{a}^{3}(x) \Biggr] \Biggr) \ ,
\ee

It is clear that if the saddle-point solution is unique, then from the
eq.(\ref{cccf}),(\ref{cccg}) one would obtain the usual RS representaion
(\ref{bbbd}),(\ref{bbbe}). However, in the case of the macroscopic
number of the local minima solutions the problem is getting highly
non-trivial. This sutuation is reminiscent of the (unsolved) problem
of summing over the saddle-point solutions in the random-field Ising model,
which is believed to provide the RSB phase near the phase
transition point \cite{rf}.

It is obviously hopeless to try to make a systematic evaluation
of the above replicated partition function.  The global solutions
$\Phi^{(\alpha)}$ are complicated implicit functions of
$\d\t(x)$.  These quantities have fluctuations of two
different types.  In the first instance, they depend on the
stochastic variables $\d\t(x)$.  But even when
the $\d\t(x)$ are completely fixed, $\Phi_{(\a)}(x)$
will depend on $\alpha$ (which labels the possible ways of
constructing the global minimum out of the choices for the
signs $\{\sigma\}$ of the local minima).  A crude way of treating
this situation is to regard the local solutions $\psi^{(i)} (x)$ as if
they were random variables, even though $\d\t(x)$ has been specified.
This randomness,
which one can see is not all that different from that which
exists in a spin glass, is the crucial one. It can be shown then, that
due to the interaction of the fluctuating fields with the
local minima configurations
(the term $\Phi_{(\a_{a})}^{2}\p_{a}^{2}$ in the eq.(\ref{cccg})),
the summation over solutions in the replica partition function
$\tl{Z_{n}}[\p_{a}]$, eq.(\ref{cccg}), could provide the additional
non-trivial RSB potential $\sum_{a,b}g_{ab}\p^{2}_{a}\p^{2}_{b}$
in which the matrix $g_{ab}$ has the Parisi RSB structure \cite{dhss}.

In the paper \cite{dhss} due to several simplifying assumptions,
the matrix $g_{ab}$ has been obtained to have explicit 1-step RSB structure,
which, in general, may not be the case.
 In this paper we are going to study the critical properties
of weakly disordered systems in terms of the RG approach
taking into account the possibility of a general type of the RSB
potentials for the fluctuating fields.
The idea is that hopefully, like in spin-glasses,
this type of generalized RG scheme self-consistently takes
into account relevant degrees of freedom coming from the
numerous local minima. In particular, the instability of the
traditional Replica Symmetric (RS) fixed points with respect to
RSB indicates that the multiplicity of the local minima can be
relevant for the critical properties in the fluctuation region.

It will be shown in the next Section that, whenever the disorder
appears to be relevant for the critical behavior,
the usual RS fixed points (which used to be considered as providing
new universal disorder-induced critical exponents) are unstable with
respect to "turning on" an RSB potential. Moreover, it will be shown that
in the presence of a general type of the RSB potentials the RG
flows actually lead to the {\it strong coupling regime} at the finite
spatial scale $R_{*} \sim \exp(1/u)$ (which corresponds to the temperature
scale $\t_{*} \sim \exp(-\fr{1}{u})$). At this scale the renormalized
matrix $g_{ab}$ develops strong RSB, and the values of the interaction
parameters are getting non-small.

Usually the strong coupling situation indicates that sertain essentially
non-perturbative exitations have to be taken into account,
and it could be argued that in the present model these are
due to exponentially rare "instantons" in the spatial regions,
where the value of $\d\t(x) \sim 1$, and the local value
of the field $\p(x)$ must be $\sim \pm 1$. (Distant analog of this
situation exists in the 2D Heisenberg model where the Poliakov
renormalization develops into the stong coupling regime at a finite
(exponentially large) scale which is known to be due to the non-linear
localized instanton solutions \cite{polyakov}).

In Section 3 we discuss the physical concequences of the obtained RG
solutions. In partiqular we show that due to the absence of fixed points
at the disorder dominated scales $R >> u^{-\nu/\a}$
(or at the corresponding temperature scales $\t << u^{1/\a}$)
there must be no simple scaling of the correlation functions
or of other physical quantities. Besides, we demonstrate, that the
structure of the SG type two-points correlation functions
is characterized by the strong RSB, indicating on the onset
of a new type of the critical behaviour of the SG nature.

In Section 4  we consider the special case of systems with the number of
spin components $p=4$, in which the pure system specific heat critical
exponent $\a = 0$. Here the disorder appears to be marginally irrelevant
in a sense that it does not change the critical exponents. Nevertheless,
the critical behaviour itself (described in terms of the logarithmic
singularities) is effected by the disorder, and moreover, the RSB phenomena
is demonstrated to be relevant in this case as well.

The remaining problems as well as future perspectives
are discussed in the Conclusions. In particular, we discuss
possible relevance of the considered RSB phenomena for the
Griffith phase which is known to exist in a finite temperature
interval near $\T$ \cite{griff}.

\sectio{Replica Symmetry Breaking in the Renormalization Group Theory}

We consider the $p$-component ferromagnet with quenched random
effective temperature fluctuations, which near the transition point
can be described by the usual Ginzburg-Landau Hamiltonian:
\be
\label{bbba}
\begin{array} {l}
H[\d\t,\f] = \I \Biggl[
\2\sum_{i=1}^{p}(\n\f_{i}(x))^{2} \\
\\
+ \2(\t - \d\t(x))\sum_{i=1}^{p}\f_{i}^{2}(x) +
\4 g\sum_{i,j=1}^{p}\f_{i}^{2}(x)\f_{j}^{2}(x) \Biggr] \ ,
\end{array}
\ee
where the quenched random temperature $\d\t(x)$ is described by the
Gaussian distribution (\ref{aaab}).

In terms of the standard replica approach
after integration over $\d\t(x)$  for the replica partition function one gets:

\be
\label{bbbdd}
\begin{array}{l}
Z_{n} \equiv \ol{Z^{n}(\d\t)} = \\
\\
\int D\f_{i}^{a}(x) \exp \Biggl[ - \I \Biggl(
\2\sum_{i=1}^{p}\sum_{a=1}^{n}[\n\f_{i}^{a}(x)]^{2}
+ \2\t\sum_{i=1}^{p}\sum_{a=1}^{n}[\f_{i}^{a}(x)]^{2} \\
\\
+ \4 \sum_{i,j=1}^{p}\sum_{a,b=1}^{n} g_{ab}
[\f_{i}^{a}(x)]^{2}[\f_{j}^{b}(x)]^{2} \Biggr) \Biggr] \ ,
\end{array}
\ee
where
\be
\lb{bbbee}
g_{ab} = g\d_{ab} - u \ .
\ee

To study the critical properties of this system we use
the standard RG procedure developed for dimensions
$D = 4 - \e$, where $\e \ll 1$.  Along the lines of the usual
rescaling scheme (see e.g. \cite{rg}) one gets the following (one-loop)
RG equations for the interaction parameters $g_{ab}$:

\be
\lb{bbbff}
\fr{d g_{ab}}{d\xi} = \e g_{ab} - \fr{1}{8\pi^{2}}
(4g_{ab}^{2} + 2(g_{aa}+g_{bb})g_{ab} + p \sum_{c=1}^{n} g_{ac} g_{cb}) \ ,
\ee
where $\xi$ is the standard rescaling parameter.

Changing $g_{ab} \to 8\pi^{2} g_{ab}$,
and $g_{a\not= b} \to -g_{a\not= b}$ (so that the off-diagonal
elements would be positively defined), and introducing $\tl{g} \equiv g_{aa}$,
we get the following RG equations:

\be
\lb{bbbf}
\fr{d g_{ab}}{d\xi} = \e g_{ab} -
(4 + 2p)\tl{g}g_{ab} + 4g^{2}_{ab} + p \sum_{c\not= a,b}^{n} g_{ac} g_{cb}
\; \; \; \; \; \; \; (a \not= b) ,
\ee
\be
\lb{bbbfff}
\fr{d}{d\xi}\tl{g} = \e\tl{g} - (8+p)\tl{g}^{2} -
p\sum_{c\not= 1}^{n} g^{2}_{1c}
\ee

If one takes the matrix $g_{ab}$ to be replica symmetric,
as in the starting form of Eq. (\ref{bbbee}), then one would
recover the usual RG equations
for the parameters $g$ and $u$, and eventually one would obtain
the well known results for the fixed points and the critical
exponents \cite{new,newnew}.
Here we leave apart the question as to how perturbations out of
the RS subspace could arise (see discussion in \cite{dhss}) and formally
consider the RG eqs.(\ref{bbbf}),(\ref{bbbfff}) assuming that the matrix
$g_{ab}$ has a general Parisi RSB structure.

According to the standard technique of the Parisi RSB algebra
(see e.g. \cite{sg}), in the limit $n\to 0$ the
matrix $g_{ab}$ is parametrized in terms of its diagonal
elements $\tl g$ and the off-diagonal {\it function} $g(x)$
defined in the interval $0<x<1$. All the operations with the matrices in
this algebra can be performed according to the following simple rules
(see e.g. \cite{manifold},\cite{intro}):

\be
\lb{bbbh}
g^{k}_{ab} \to (\tl g^{k}; g^{k}(x)) ,
\ee
\be
\lb{bbbi}
(\hat g^{2} )_{ab} \equiv  \sum_{c=1}^{n} g_{ac} g_{cb} \to (\tl c; c(x)) ,
\ee
where
\be
\lb{bbbj}
\ba
\tl c = \tl g^{2} - \int_{0}^{1} dx g^{2}(x) ,\\
\\
c(x) = 2(\tl g - \int_{0}^{1} dy g(y) ) g(x) -
\int_{0}^{x} dy [g(x) - g(y)]^{2} .
\ea
\ee
The RS situation corresponds to the case $g(x) = const$ independent of $x$.

Using the above rules from the eqs.(\ref{bbbf}),(\ref{bbbfff}) one gets:

\be
\lb{aabb}
\fr{d}{d\xi}g(x) = (\e - (4+2p)\tl{g})g(x) + 4g^{2}(x)
-2pg(x)\int_{0}^{1}dy g(y) - p\int_{0}^{x} dy (g(x)-g(y))^{2}
\ee
\be
\lb {aacc}
\fr{d}{d\xi}\tl{g} = \e\tl{g} - (8+p)\tl{g}^{2} + p\ol{g^{2}}
\ee
where $\ol{g^{2}} \equiv \int_{0}^{1}dx g^{2}(x)$.

Usually in the studies of the critical behaviour one is looking for the
stable fixed-points solutions of the RG equations. The fixed-point values
of the of the renormalized interaction parameters are believed to
describe the asymptotic structure of the effective Hamiltonian
which makes possible to calculate the singular part of the free energy, as
well as the other thermodynamic quantities.

{}From the eq.(\ref{aabb}) one can easily find out what should
be the structure of the function $g(x)$ at the fixed point,
$\fr{d}{d\xi}g(x) = 0$, $\fr{d}{d\xi}\tl{g} = 0$.
Taking the derivative over $x$ twice, one gets, from Eq.
(\ref{aabb}): $g'(x) = 0$. This means that either the
function $g(x)$ is constant (which is the RS situation), or it
has the step-like structure. It is interesting to note that the structure
of fixed-point equations is similar to those for the Parisi
function $q(x)$ near $\T$ in the Potts spin-glasses \cite{potts}, and it is
the term $g^{2}(x)$ in Eq.(\ref{aabb})  which is known to produce 1step RSB
solution there. The numerical solution of the above RG equations
convincingly demonstrates that whenever the triel function $g(x)$
has the many-step RSB structure, it quickly developes into the 1-step
one with the coordinate of the step being the most right step
of the original many-step function.

Let us consider the 1-step RSB ansatz for the function $g(x)$:

\be
\lb{bbbm}
g(x) = \l\{ \begin{array}{ll}
g_{0}        & \mbox{for $0 \leq x < x_{0}$}  \\
g_{1}        & \mbox{for $x_{0} < x \leq 1$}
\end{array}
\right.
\ee
where $0 \leq x_{0} \leq 1$ is the coordinate of the step.

In terms of this ansatz the above fixed-point equations have
several non-trivial solutions:

\vspace{3mm}

1) The RS fixed-point which corresponds to the pure system:

\be
\lb{bbbo}
g_{0} = g_{1} = 0; \; \; \; \tl{g} = \fr{1}{8+p}\e
\ee
This fixed point (in accordance with the Harris criterion) is
stable for the number of spin components $p > 4$, and it is getting unstable
for $p < 4$.

\vspace{3mm}

2) The disorder-induced RS fixed point (for $p > 1$)\cite{new,newnew}:

\be
\lb{bbbp}
g_{0} = g_{1} = \e\fr{4-p}{16(p-1)}; \; \; \; \tl{g} = \e\fr{p}{16(p-1)} .
\ee
It was usually considered to be the one which describes the new universal
critical behaviour in systems with impurities. This fixed point has
been shown to be stable (with respect to the RS deviations!)
for $p < 4$, which is consistent with the Harris criterion.
(For $p=1$ this fixed point involves an expansion in powers of
$( \epsilon )^{1/2}$ and this structure is only revealed
within a two-loop approximation). However, the stability analysis with respect
to the RSB deviations shows that this fixed point is {\it always unstable}
\cite{dhss}. Therefore, whenever the disorder is relevant for the
critical behaviour, the RSB perturbations must be getting the dominant factor
in the asymptotic large scale limit.

\vspace{3mm}

3) The 1-step RSB fixed point \cite{dhss}:

\be
\lb{bbbr}
\ba
g_{0} = 0 ; \; \; \;
g_{1} = \e\fr{4-p}{16(p-1) - px_{0}(8+p)} ,\\
\\
\tl{g} = \e\fr{p(1-x_{0})}{16(p-1) - px_{0}(8+p)} .
\ea
\ee
This fixed point can be shown to be stable (within 1-step RSB subspace!)
for:

\be
\lb{bbbrr}
\ba
1 < p < 4 ,\\
\\
0 < x_{0} < x_{c}(p) \equiv \fr{16(p-1)}{p(8+p)} .
\ea
\ee
In particular, $x_{c}(p=2) = 4/5$; $x_{c}(p=3) = 32/33$, and $x_{c}(p=4) = 1$.
Using the result (\ref{bbbr}) one can easily obtain the corresponding
critical exponents which are now getting to be non-universal
being dependent on the starting parameter $x_{0}$ \cite{dhss} (see also
next Section).

(Note, that in addition to the fixed points listed above there exist several
other 1step RSB solutions which are either unstable or unphysical.)

\vspace{3mm}

The problem, however, is that if the parameter $x_{0}$ of the starting
function $g(x;\xi=0)$ (or, more generaly, the coordinate of the most right
step of the many-steps starting function) is beyond the stability interval,
such that $x_{c}(p) < x_{0} < 1$, then there exist {\it no stable
fixed points} of the RG eqs.(\ref{aabb}),(\ref{aacc}). One faces
the same situation, of course, in the case of a general continuous
starting function $g(x;\xi=0)$. Moreover, according to eq.(\ref{bbbrr})
there exist no stable fixed points out of the RS subspace in the most
intersting Ising case, $p=1$.

Unlike the RS situation for $p=1$, where one finds the
stable $\sim \sqrt{\e}$ fixed point in the two-loop RG equations \cite{newnew},
here adding next order terms in the RG equations doesn't cure the problem.
In the considered RSB case one finds that in the two-loops
RG equations the values of the parameters in the fixed point
are formally getting of the order of one, and it signals that we are
entering the strong coupling regime where all the orders of the RG are
getting relevant.

Nevertheless, to get at least some information about the physics
behind this instability phenomena, one can proceed analizing the
actual evolution of the above one-loop RG equations. The scale evolution
of the parametrs of the Hamiltonian would still adequately describe the
properties of the system until we reach a critical scale $\xi_{*}$, at which
the strong coupling regime begins.

The evolution of the renormalized function $g(x;\xi)$ can be analyzed
both numerically and analytically.
It can be shown (see Appendix A) that in the case $p < 4$ for
a general continuous starting function $g(x;\xi=0) \equiv g_{0}(x)$
the renormalized function $g(x;\xi)$ tends to zero everywhere
in the interval $0 \leq x < (1-\D(\xi))$, while in the narrow
(scale dependent) interval $\D(\xi)$ near $x=1$ the values of
of the function $g(x;\xi)$ grow:

\be
\lb{aann}
g(x;\xi) \sim \left\{ \begin{array}{ll}
a\fr{u}{1 - u\xi} ; \; \; \; \mbox{at $(1-x) << \D(\xi)$}\\
\\
0 ; \; \; \;  \mbox{at $(1-x) >> \D(\xi)$}
\end{array}
\right.
\ee
\be
\lb{aannn}
\tl{g}(\xi) \sim u \ln\fr{1}{1-u\xi}
\ee
where
\be
\lb{aannnn}
\D(\xi) \simeq (1 - u\xi)
\ee
Here $a$ is a positive non-universal constant, and the
critical scale $\xi_{*}$ is defined by the condition that
the values of the renormalized parameters are getting of the order of one:
$(1 - u\xi_{*}) \sim u$, or $\xi_{*} \sim 1/u$. Correspondingly,
the spatial scale at which the system is entering the strong coupling
regime is:

\be
\lb{aaaaz}
R_{*} \sim \exp(\fr{1}{u})
\ee
Note that the value of this scale is much bigger than the usual crossover
scale $\sim u^{-\a/\nu}$ (where $\a$ and $\nu$ are the
pure system specific heat and the correlation length critical exponents),
at which the disorder is getting relevant for the critical behaviour.

According to the above result, the value of the narrow band near
$x=1$ where the function $g(x;\xi)$ is formally getting
divergent is $\D(\xi) \simeq (1 - u\xi) \to u << 1$ as $\xi \to \xi_{*}$.

Besides, it can also be shown (Appendix A) that the value of the integral
$\ol{g}(\xi) \equiv \int_{0}^{1} g(x;\xi)$
is formally getting divergent logarithmically as $\xi \to \xi_{*}$:

\be
\lb{aaaaa}
\ol{g}(\xi) \sim u\ln\fr{1}{1-u\xi}
\ee

Qualitatively similar asymptotic behaviour for $g(x;\xi)$
is obtained for the case when the starting function $g_{0}(x)$ has the
1-step RSB structure (\ref{bbbm}), and the coordinate of the
step $x_{0}$ is in the instability region (or for any $x_{0}$ in the Ising
case $p=1$):

\be
\lb{aaabz}
g(x;\xi) \sim \left\{ \begin{array}{ll}
\fr{g_{1}(0)}{1 - (4-2p+px_{0})g_{1}(0)\xi} ;
\; \mbox{at $x_{0} < x < 1$}\\
\\
0 ; \; \; \;  \mbox{at $0 \leq x < x_{0}$}
\end{array}
\right.
\ee
Here $g_{1}(0) \equiv g_{1}(\xi=0) \sim u$, and the coefficient
$(4-2p+px_{0})$ is always positive. In this case again, the system
arrives into the strong coupling regime at scales $\xi \sim 1/u$.

Note that the above asymptotics do not explicitely involve $\e$.
Actually, the role of the
parameter $\e > 0$ is to "push" the RG trajectories out of the
trivial Gaussian fixed point $g = 0; \tl{g} = 0$. Thus, the value
of $\e$, as well as the values of the starting parameters
$g_{0}(x)$, $\tl{g}_{0}$, define a scale at which the solutions
finally arrive to the above asymptotic regime. In the case
$\e < 0$ (above dimensions 4) the Gaussian fixed point is stable;
on the other hand, the strong coupling asymptotics still exists
in this case as well, separated from the trivial one by a finite
(depending on the value of $\e$) barrier. Therefore, although
{\it infinitely small} disorder remains irrelevant for the critical
behaviour above dimensions 4, if the disorder is strong  enough
(bigger than sertain depending on $\e$ threshould value) the RG
trajectories could arrive to the above strong coupling regime again.

\sectio{Scaling and Correlation functions}

\subsection{Temperature Scales}

The renormalization of the mass term $\t(\xi)\sum_{a=1}^{n}\f_{a}^{2}$
is described by the following RG equation:

\be
\lb{cca}
\fr{d}{d\xi}\ln\t = 2 - \fr{1}{8\pi^{2}}
[(2+p)\tl{g} + p\sum_{a\not= 1}^{n} g_{1a}]
\ee
Changing (as in the previous section) $g_{ab} \to 8\pi^{2} g_{ab}$,
and $g_{a\not= b} \to -g_{a\not= b}$, in the Parisi representation we get:

\be
\lb{ccb}
\fr{d}{d\xi}\ln\t = 2 - [(2+p)\tl{g}(\xi) + p\int_{0}^{1} g(x;\xi)]
\ee
or

\be
\lb{ccc}
\t(\xi) = \t_{0}\exp\{2\xi -
\int_{0}^{\xi}d\eta [(2+p)\tl{g}(\eta) + p\ol{g}(\eta)] \}
\ee
where $\tl{g}(\eta)$ and $\ol{g}(\eta) \equiv \int_{0}^{1}dx g(x;\eta)$
are the solutions of the RG equations of the previous section.

\vspace{3mm}

Consider first what was the traditional (replica-symmetric) situation.
The RS interaction parameters $\tl{g}(\xi)$ and $g(\xi)$ are arriving
to the fixed point values $\tl{g}_{*}$ and $g_{*}$ (which are of the order of
$\e$), and then for the dependence of the renormalized mass $\t(\xi)$,
according to (\ref{ccc}), one gets:

\be
\lb{ccd}
\t(\xi) = \t_{0}\exp\{ \D_{\t}\xi \}
\ee
where

\be
\lb{cce}
\D_{\t} = 2 - [(2+p)\tl{g}_{*} + p g_{*}]
\ee
At scale $\xi_{c}$, such that $\t(\xi_{c})$ is getting of the order of one,
the system gets out of the scaling region. Since the RG parameter
is by definition $\xi = \ln R$, where $R$ is the spartial scale,
this defines the correlation length $\R$ as a function of the reduced
temperature $\t_{0}$. According to (\ref{ccd}), one obtaines:

\be
\lb{ccf}
\R(\t_{0}) \sim \t_{0}^{-\nu}
\ee
where $\nu = 1/\D_{\t}$, eq.(\ref{cce}), is the critical exponent of the
correlation length.

Actually, if the starting value of the disorder parameter $g(\xi=0) \equiv u$
is much smaller than starting value of the pure system interaction
$\tl{g}(\xi=0) \equiv g_{0}$, the sutuation is a little bit more
complicated. In this case the RG flow for $\tl{g}(\xi)$ first arrives
to the pure system fixed point $\tl{g}_{*}^{(pure)}$, as if the disorder
perturbation does not exist. Then, since the pure system fixed point is
unstable with respect to the disorder perturbations, at scales bigger than
sertain disorder dependent scale $\xi_{u}$ the RG trajectories
are eventually arriving to the stable (universal) disorder induced
fixed point ($\tl{g}_{*}$, $g_{*}$). According to the traditional theory
\cite{new} it is known that $\xi_{u} \sim \fr{\nu}{\a}\ln\fr{1}{u}$.
The corresponding spartial
scale is $R_{u} \sim u^{-\nu/\a}$, and it is big it terms of the
small parameter $u$.

Coming back to the scaling behaviour of the mass parameter $\t(\xi)$,
eq.(\ref{ccd}), we see that if the value of the temperature $\t_{0}$
is such that $\t(\xi)$ is getting of the order of one before the
crossover scale $\xi_{u}$ is reached, then for the scaling behaviour
of the correlation length (as well as for other thermodynamic quatities)
one finds essentially the pure system result
$\R(\t_{0}) \sim \t_{0}^{-\nu_{(pure)}}$. However, the pure system
critical behaviour is observed only untile $\R << R_{u}$, which imposes
the restriction on the temperature parameter:
$\t_{0} >> u^{1/\a} \equiv \t_{u}$. In other words, at temperatures
not too close to $\T$, $\t_{u} << \t_{0} << 1$, the presence of disorder
is irrelevant for the critical behaviour.

On the other hand, if $\t_{0} << \t_{u}$ (in the close vicinity of $\T$),
the RG trajectories for $\tl{g}(\xi)$ and $g(\xi)$ are arriving
(after crossover) into a new (universal) disorder induced fixed point
($\tl{g}_{*}$, $g_{*}$), and the scaling of the correlation length
(as well as other thermodynamic quantities),
according to eqs.(\ref{ccf})-(\ref{cce}), is getting to be controled
by a new universal critical exponent $\nu$ which is defined by the RS
fixed point ($\tl{g}_{*}$, $g_{*}$) of the random system.

\vspace{3mm}

Consider now what is the situation if the RSB scenario takes place.
Again, if the disorder parameter $u$ is small, in the temperature interval
$\t_{u} << \t_{0} << 1$, the critical behaviour is essentially
controlled by the pure system fixed point, and the presence of disorder
is irrelevant. For the same reasons as discussed above, the system
gets out of the scaling regime ($\t(\xi)$ is getting of the order of one)
before the disorder parameters start "pushing" the RG trajectories
out of the pure system fixed point.

However, at temperatures $\t_{0} << \t_{u}$ the situation is getting
completely different from the RS case.
At scales $\xi >> \xi_{u}$ (although still $\xi << \xi_{*} \sim \fr{1}{u}$)
according to the solutions (\ref{aann}), (\ref{aaabz})
the parameters $\tl{g}(\xi)$ and $g(x;\xi)$,
does not arrive to any fixed point, and they keep evolving as the
scale $\xi$ increases. Therefore, here, according to eq.(\ref{ccc}),
the correlation length
(defined, as usual, by the condition that the renormalized
$\t(\xi)$ is getting of the order of one) is getting to be defined
by the following non-trivial equation:

\be
\lb{ccg}
2 \ln\R - \int_{0}^{\ln\R}d\eta [(2+p)\tl{g}(\eta) + p\ol{g}(\eta)] =
\ln\fr{1}{\t_{0}}
\ee
Thus, as the temperature is getting sufficiently close to $\T$
(in the disorder dominated region $\t_{0} << \t_{u}$) there will be no usual
scaling dependence of the
correlation length (as well as other thermodynamic quantities)
like in the eq.(\ref{ccf}).

Finally, as the temperature parameter $\t_{0}$ is getting smaller and
smaller, what happens is that at scale
$\xi_{*} \equiv \ln R_{*} \sim \fr{1}{u}$
we are entering into the strong coupling regime (such that the
parameters $\tl{g}(\xi)$ and $g(x;\xi)$ are getting non-small), while
the renormalized mass $\t(\xi)$ remains still small.

According to the solution obtained in Appendix A, the
integrals $\int_{0}^{\xi_{*}}d\eta \tl{g}(\eta)$ and
$\int_{0}^{\xi_{*}}d\eta\ol{g}(\eta) \equiv G_{c}$ have a finite
(depending on the initial conditions) value. Thus, according to
eq.(\ref{ccg}), for the crossover temperature we get:

\be
\lb{cch}
\t_{*} \sim \exp(-\fr{const}{u})
\ee

\vspace{5mm}

In the close vicinity of $\T$ at $\t << \t_{*}$
we are facing the situation that at large scales the interaction
parameters of the asymptotic (zero-mass) Hamiltonian are getting
non-small, and
the properties of the system can not be anylized in terms of simple
one-loop RG approach. Nevertheless, the
qualitative structure of the asymptotic Hamiltonian makes it possible
to argue that in the temperature interval $\t << \t_{*}$ near $\T$
the properties of the system should be essentially SG-like.
The point is that it is the parameter describing the disorder,
$g(x;\xi)$, which is the most divergent.

In a sense, here the problem is qualitetively reduced back to the
original one with {\it strong} disorder at the critical point.
It doesn't seem probable, however, that the state of the system
will be described by non-zero true SG order parameter
$Q_{ab} = \la\f_{a}\f_{b}\ra$ (which would mean real SG freezing).
Otherwise there must exist finite value of $\t$ at which
real thermodynamic phase transition into the SG phase takes place,
while we observe only the {\it crossover} temperature $\t_{*}$,
at which change of critical regime occurs.

It seems more realistic to expect that at scales $\sim \xi_{*}$
the RG trajectories finally arrive to a fixed-point characterized
by non-small values of the interaction parameters and strong RSB.
Then, the SG-like  behaviour of the system near $\T$ will be characterized
by its highly non-trivial critical properties exhibiting
strong RSB phenomena.

\subsection{Correlation Functions}

Consider the scaling properties of the spin-glass type
connected correlation function:

\be
\lb{dda}
K(R) = \ol{(\la\f(0)\f(R)\ra - \la\f(0)\ra\la\f(R)\ra)^{2}}
\equiv \ol{\la\la\f(0)\f(R)\ra\ra^{2}}
\ee
In terms of the replica formalism one gets:

\be
\lb{ddb}
K(R) = \lim_{n\to 0} \fr{1}{n(n-1)}\sum_{a\not=b}^{n} K_{ab}(R)
\ee
where

\be
\lb{ddc}
K_{ab}(R) = \la\la\f_{a}(0)\f_{b}(0)\f_{a}(R)\f_{b}(R)\ra\ra
\ee

In terms of the standard RG formalizm for the replica correlation function
$K_{ab}(R)$ one finds:

\be
\lb{ddd}
K_{ab}(R) \sim (G_{0}(R))^{2} (Z_{ab}(R))^{2}
\ee
where

\be
\lb{ddf}
G_{0}(R) = R^{-(D-2)}
\ee
is the free-field correlation function, and in the one-loop approximation
the scaling of the mass-like object $Z_{ab}(R)$ (with $a\not= b$)
is defined by the RG equation:

\be
\lb{ddg}
\fr{d}{d\xi}\ln Z_{ab}(\xi) =  2 g_{ab}(\xi)
\ee
Here $g_{a\not= b}(\xi) > 0$ is the solution of the corrsponding RG equations
(\ref{bbbf})-(\ref{bbbfff}), $\xi = \ln R$, and $Z_{ab}(0) \equiv 1$.

For the correlation function (\ref{ddd}) one finds:

\be
\lb{ddh}
K_{ab}(R) \sim (G_{0}(R))^{2}
\exp\{ 4 \int_{0}^{\ln R} d\xi g_{ab}(\xi) \}
\ee
Correspondingly, in the Parisi representation:
$g_{a\not= b}(\xi) \to g(x;\xi)$ and $K_{a\not= b}(R) \to K(x; R)$,
one gets:

\be
\lb{ddhh}
K(x; R) \sim (G_{0}(R))^{2}
\exp\{ 4 \int_{0}^{\ln R} d\xi g(x;\xi) \}
\ee

To realize the effects of the RSB more clearly consider again what was the
situation in the traditional RS case. Here (for $p < 4$) one finds that
the interaction
parameter $g_{a\not= b}(\xi) \equiv u(\xi)$ arrives to the RS fixed
point $u_{*} = \e \fr{4-p}{16(p-1)}$, and according to
eqs.(\ref{ddh}),(\ref{ddb}) one obtains simple scaling:

\be
\lb{ddii}
K_{rs}(R) \sim R^{-2(D - 2) + \theta}
\ee
with the universal disorder induced critical exponent

\be
\lb{ddj}
\theta = \e\fr{4-p}{4(p-1)}
\ee

In the case of the 1-step RSB fixed point, eq.(\ref{bbbr}),
the situation is getting somewhat more complicated. Here
one finds that the correlation function $K(x; R)$ also have 1RSB
structure:

\be
\lb{ddi}
K(x;R) \sim  \left\{ \begin{array}{ll}
                   K_{0}(R); \; \; \; \; \; \mbox{for $0 \leq x < x_{0}$}\\
                   K_{1}(R); \; \; \; \; \; \mbox{for $x_{0} < x \leq 1$}\\
                 \end{array}
                 \right.
\ee
where (in the first order in $\e$)

\be
\lb{ddjj}
\ba
K_{0}(R) \sim R^{-2(D - 2)} = G_{0}^{2}(R) \\
\\
K_{1}(R) \sim R^{-2(D - 2) + \theta_{1rsb}}
\ea
\ee
with {\it non-universal} critical exponent $\theta_{1rsb}$ explicitely
depending on the coordinate of the step $x_{0}$:

\be
\lb{ddk}
\theta_{1rsb} = \e\fr{4(4-p)}{16(p-1) - px_{0}(8+p)}
\ee

Since the critical exponent $\theta_{1rsb}$ is positive, the leading
contribution to the "observable" quantity
$K(R) = \ol{\la\la\f(0)\f(R)\ra\ra^{2}}$, eq.(\ref{ddb}), is given
by $K_{1}(R)$:

\be
\lb{ddl}
K(R) \sim (1-x_{0})K_{1}(R) + x_{0}K_{0}(R) \sim
R^{-2(D - 2) + \theta_{1rsb}}
\ee

But the difference between the 1RSB the RS cases must be observed
not only in the result that
their critical exponents $\theta$ of the correlation functions
$K(R)$ must be different. According to the traditional SG philosophy \cite{sg},
the result that the scaling of the
RSB correlation function $K_{ab}(R)$ or $K(x;R)$ does depend on the
replica indices $(a,b)$ or the replica parameter $x$, eq.(\ref{ddi}),
indicates that
in different measurements of the correlation function for {\it the same}
realization of the quenched disorder one is going to obtain
{\it different} results, $K_{0}(R)$ or $K_{1}(R)$, with the probabilities
defined by the value of $x_{0}$.

In real experiments, however, one is dealing with the quantities
averaged in space. In particular, for the two-point correlation functions
the measurable quantity is obtained by integration over the two points,
such that the distance $R$ between them is fixed. Of course, the result
obtained this way must be equivalent simply to $K(R)$, eq.(\ref{ddl}),
found by formal everaging over different realizations of disorder,
and different scalings  $K_{0}(R)$ and $K_{1}(R)$ can not be observed
this way.

Nevertheless, for somewhat different scheme of the
measurements the qualitative difference with the RS situation can
be observed. In spin-glasses it is generally believed that RSB can be
interpreted as factorization of the phase space into (ultrametric)
hierarchy of "valleys", or local minima pure states separated by
macroscopic barries. Although in the present case the local minima
configurations responcible for the RSB can not be separated by
infinite barriers, it would be natural to interpret obtained phenomenon
as effective factorization of the phase space into a hierarchy
of valleys separated by {\it finite} barriers. Since the only
relevant scale in the critical region is the correlation length
the maximum energy barriers must be proportional to $\R^{D}(\t)$,
and they are getting divergent as the critical temperature is approached.
In this situation one could expect that besides the usual critical slowing
down (corresponding to the relaxation inside one
valley) qualitatively much bigger relaxation times would be required
for overcoming barriers separated different valleys. Therefore,
the traditional measuremets of the observables in the "thermal equilibrium"
can actually correspond to the equlibration within one valley only
and not to the true thermal equilibrium. Then in different measuremets
(for the same sample) one could be effectively "trapped" in different
valleys and thus the traditional spin-glass situation is restored.

To check whether the above speculations are correct or not, like in
spin-glasses one can invent traditional "overlap" quantities
which could hopefully reveal the existance of the multiple valley
structures. For instance, one can introduce the spartially averaged
quantity for {\it pairs} of different realizations of the disorder:

\be
\lb{ddm}
K_{ij}(R) \equiv \fr{1}{V}\int d^{D}r
\la\f(r)\f(r+R)\ra_{i}\la\f(r)\f(r+R)\ra_{j}
\ee
where $i$ and $j$ label different realizations, and it is assumed
that the measurable thermal average corresponds to a particular valley,
and not to the true thermal average. If the RS situation
takes place (so that only one global valley exists), then for different
pairs of realizations one will be obtaining
the same result (\ref{ddii}). On the other hand, in the case of the 1RSB,
according to the general theory of the RSB \cite{sg},
after obtaining statistics over pairs of realizations for $K_{ij}(R)$
one has to be getting the result $K_{0}(R)$ with the probability
$x_{0}$, and $K_{1}(R)$ with the probability $(1-x_{0})$.

\vspace{5mm}

Consider finally what would be the situation if a general type
of the RSB takes place.
According to the qualitative solution (\ref{aann})-(\ref{aannn}), the
function $g(x;\xi)$ does not arrive to any fixed point at scales
$\xi >> \xi_{u} \sim \fr{\nu}{\a}\ln\fr{1}{u}$. Therefore,
at the disorder dominanted scales $R >> R_{u} \sim u^{-\nu/\a} >>1$
there must be no scaling behaviour of the correlation function $K(R)$.
Near the critical scale $\xi_{*} \sim 1/u$ the qualitative behaviour
of the solution $g(x;\xi)$ is shown in eq.(\ref{aann}).
Therefore, according to eq.(\ref{ddhh}), near the critical scale
$R_{*} \sim \exp(1/u)$ for the correlation function
$K(x;R)$ one obtaines:

\be
\lb{ddn}
K(x;R) \sim  \left\{ \begin{array}{ll}
R^{-2(D-2)} (1 - u\ln R)^{-4a} \equiv K_{1}(R); \; \; \;
\mbox{for $(1-x)<<\D(R)$}\\
\\
R^{-2(D-2)} = G_{0}^{2}(R) \equiv K_{0};  \; \; \; \mbox{for $(1-x)>>\D(R)$}
\end{array}
\right.
\ee
where $\D(R) = (1 - u\ln R) \to u << 1$ as $R \to R_{*}$.

At the critical scale
one has $(1 - u\ln R_{*}) \sim u $, and according to eq.(\ref{ddn})
the shape of the replica function $K(x;R)$ must be "quasi-1step":

\be
\lb{ddo}
K(x;R_{*}) \sim  \left\{ \begin{array}{ll}
u^{-4a} \exp\{-\fr{2(D-2)}{u}\} \equiv K_{1}^{*};
\; \; \mbox{for $(1-x) << u$}\\
\\
\exp\{-\fr{2(D-2)}{u}\} \equiv K_{0}^{*}; \; \; \; \; \mbox{for $(1-x) >> u$}
\end{array}
\right.
\ee

According to the above discussion of the observale quantities for the
1-step RSB case, the result (\ref{ddo}) could be measured for the
spartially averaged overlaps of the correlation functions $K_{ij}(R)$,
eq.(\ref{ddm}), for the statistics of pairs of realizations of the
disorder. Then, for the correlation function $K_{ij}(R)$ one
is expected to be obtaining the value $K_{1}$ with the small probability $u$
and the value $K_{0}$ with the probability $(1-u)$. Although both
values $K_{1}^{*}$ and $K_{0}^{*}$ are expected to be exponentially
small, their ratio $K_{1}^{*}/K_{0}^{*} \sim u^{-4a}$ must be big.

Finally, at scales $R >> R_{*}$ we are entering into the strong coupling
regime, where simple one-loop RG approach can not be used any more.

\subsection{Specific Heat}

According to the standard procedure the leading singularity of the
specific heat can be calculated as follows:

\be
\lb{eea}
C \sim \int d^{D}R [\ol{\la\f^{2}(0)\f^{2}(R)\ra} -
\ol{\la\f^{2}(0)\ra\la\f^{2}(R)\ra}]
\ee
In terms of the RG scheme for the correlation function:

\be
\lb{eeb}
W(R) \equiv \ol{\la\f^{2}(0)\f^{2}(R)\ra} -
\ol{\la\f^{2}(0)\ra\la\f^{2}(R)\ra}
\ee
one gets:

\be
\lb{eec}
W(R) = (G_{0}(R))^{2} m^{2}(R)
\ee
where $G_{0}(R) = R^{-(D-2)}$ is the free field two-point correlation function,
and the mass-like object $m(R)$ is given by the solution
of the following (one-loop) RG equation (c.f. eq.(\ref{ccb})):

\be
\lb{eed}
\fr{d}{d\xi} \ln m(\xi) =
 -[(2+p)\tl{g}(\xi) - p\sum_{a\not= 1}^{n} g_{a1}(\xi)]
\ee
Here, as usual, $\xi = \ln R$, and the renormalized interaction parameters
$\tl{g}(\xi)$ and $g_{a\not= b}(\xi)$ are the solutions of the replica
RG equations (\ref{bbbf})-(\ref{bbbfff}). In the Parisi representation,
$g_{a\not= b}(\xi) \to g(x;\xi)$, one gets:

\be
\lb{eee}
m(R) = \exp\{
 -(2+p)\int_{0}^{\ln R} d\xi\tl{g}(\xi)
 - p\int_{0}^{\ln R} d\xi \int_{0}^{1} dx g(x;\xi) \}
\ee
Then, after simple transformations for the singular part of the
specific heat, eq.(\ref{eea}), one gets:

\be
\lb{eef}
C \sim \int_{0}^{\xi_{max}} d\xi
\exp\{ \e\xi
 -2(2+p)\int_{0}^{\xi} d\eta\tl{g}(\eta)
 - 2p\int_{0}^{\xi} d\eta \ol{g}(\eta) \}
\ee
where $\ol{g}(\eta) \equiv \int_{0}^{1} dx g(x;\eta)$.
The infrared cut-off $\xi_{max}$ in (\ref{eef}) is the scale at which
the system get out of the scaling regime.

Usually $\xi_{max}$ is the scale at which the renormalized mass
$\t(\xi)$, eq.(\ref{ccc}), is getting of the order of one, and
if the traditional scaling situation takes place, one finds that
$\xi_{max} \sim \ln(1/\t_{0})$.

Again, consider first what was the situation in the traditional RS case.
Here at scales $\xi >> \xi_{u} \sim \ln(1/u)$ (which correspond
to the temperature region $\t_{0} << \t_{u} \sim u^{\nu/\a}$)
the renormalized parameters
$\tl{g}(\eta)$ and $g(\xi)$ are arriving into the universal fixed point
$\tl{g}_{*} = \e \fr{p}{16(p-1)}$ ; $g_{*} = \e \fr{4-p}{16(p-1)}$,
(see Section 2, eq.(\ref{bbbp}))
and according to (\ref{eef}) for the singular part of the specific heat
one finds \cite{new},\cite{newnew}:

\be
\lb{eeg}
C(\t_{0}) \sim
\int_{0}^{\ln(1/\t_{0})} d\xi
\exp\{ \xi [\e - 2(2+p)\tl{g}_{*} - 2p g_{*}] \}
\sim \t_{0}^{\e\fr{4-p}{4(p-1)}}
\ee
So that in the close vicinity of $\T$ one would expect to observe
new universal disorder induced critical behaviour with negative
specific heat critical exponent $\a = - \e\fr{4-p}{4(p-1)}$
(unlike positive $\a$ in the corresponding pure system).

Similary, if the scenario with the stable 1-step RSB fixed points
takes place, then one finds that the specific heat critical exponent
$\a(x_{0})$ is getting to be non-universal, explicitely depending
on the coordinate of the step $x_{0}$ \cite{dhss}:

\be
\lb{bbbv}
\a(x_{0}) = -\2\e\fr{(4-p)(4-px_{0})}{16(p-1)-px_{0}(p+8)} .
\ee

In the general RSB case the situation is getting completely different.
Here in the disorder dominated region $\t_{*} << \t_{0} << u^{\nu/\a}$
(which corresponds to scales $\xi_{u} << \xi << \xi_{*}$) the RG trajectories
of the interaction parameters $\tl{g}(\xi)$ and $\ol{g}(\xi)$ does not
arrive to any fixed point, and according to eq.(\ref{eeg}) one finds
that the specific heat is getting to be a complicated function of the
temperature parameter $\t_{0}$ which does not have the traditional
scaling form.

Finally, in the SG-like region in the close vicinity of $\T$, where
the interaction parameters $\tl{g}$ and $\ol{g}$ are getting finite,
one finds that the integral over $\xi$ in eq.(\ref{eef}) is getting
converging (so that the upper cut-off scale $\xi_{max}$ is getting
irrelevant). Thus, in this case one obtaines the result that
the "would be singular part" of the specific heat remains finite
in the temperature interval $\sim \t_{*}$ around $\T$, so that
the specific heat is getting {\it non-singular} at the phase transition point.

\sectio{Marginal case $p=4$}

In the systems with the number of spin components $p=4$
(in which the pure system specific heat critical exponent $\a = 0$)
the disorder
appears to be marginally irrelevant in a sense that it does not change
the critical exponents. Nevertheless, the critical behaviour
(described in terms of the logarithmic singularities) is
effected by the disorder, and moreover, the RSB phenomena appear
to relevant in this case as well.

Consider first the replica symmetric situation: $g(x;\xi) \equiv g(\xi)$.
For the RG equations (\ref{aabb}),(\ref{aacc}) one gets:

\be
\lb{zzza}
\ba
\fr{dg}{d\xi} = (\e - 12\tl{g})g - 4g^{2} \\
\\
\fr{d\tl{g}}{d\xi} = (\e - 12\tl{g})\tl{g} + 4g^{2}
\ea
\ee
In the pure system ($g \equiv 0$) the fixed point is:

\be
\lb{zzzb}
\tl{g}_{pure} = \fr{1}{12}\e
\ee
Using eq.(\ref{eeg}) for the singular part of the specific heat of
the pure system one easily finds:

\be
\lb{zzzc}
C_{pure}(\t) \sim \ln(\fr{1}{\t})
\ee
Thus, although the specific heat critical exponent of the  pure system
is zero, the specific heat is still divergent in the critical point.

For the system with disorder the (replica symmetric) asymptotic solution
of the eqs.(\ref{zzza})  is:

\be
\lb{zzzd}
\ba
g(\xi) \simeq \fr{1}{4} \xi^{-1} \to 0\\
\\
\tl{g}(\xi) \simeq \fr{1}{12}\e + q(\xi)
\ea
\ee
where
\be
q(\xi) \sim \xi^{-2} \to 0
\ee
In this case the renormalized parameters are asymptotically approaching the
pure system fixed point $\tl{g} = \e/12$, $g=0$ (so that the disorder
is marginally irrelevant). Nevertheless, due to
slow power-law approach to the fixed point the logarithmic singularity
of the specific heat is changing into another universal type.
{}From the general expression (\ref{eef}) for the singular part of the
specific heat one obtains:

\be
\lb{zzze}
C \sim \int_{0}^{\ln(1/\t)} d\xi
\exp\{ \int_{0}^{\xi} d\eta [\e -12\tl{g}(\eta)- 8g(\eta)] \}
\ee
Using the result (\ref{zzzd}) one easily finds:

\be
\lb{zzzf}
C_{rs}(\t) \sim \fr{1}{\ln(\fr{1}{\t})}
\ee
One can also easily check that (unlike the systems with $p < 4$)
the crossover from the pure system
critical behaviour, eq.(\ref{zzzc}), to the disorder induced one,
eq.(\ref{zzzf}), takes place in the exponentially small
temperature interval near $\T$:

\be
\lb{zzzg}
\t_{u} \sim \exp(-\fr{1}{u})
\ee

Consider now the effects of the RSB.
The analytic solution of the RG equations (\ref{aabb}),(\ref{aacc})
(see Appendix B) shows that there is no strong coupling regime in the
$p=4$ case, and the asymptitic behaviour (at scales $\xi >> 1/u$)
of the renormalized parameters can be found exactly:

\be
\lb{bbbz}
g(x;\xi) \sim \left\{ \begin{array}{ll}
         \xi^{-2} ,  \; \; \; \; (1-x) >> \fr{1}{\sqrt{\g\xi}}  \\
         \fr{1}{\sqrt{\g\xi}}, \; \; \; \; (1-x) << \fr{1}{\sqrt{\g\xi}} \\
         \end{array}
         \right.
\ee
\be
\lb{bbbb}
\ba
\tl{g}(\xi) \simeq \fr{\e}{12} + q(\xi) \\
\\
q(\xi) \sim \xi^{-3/2} \to 0
\ea
\ee
Here $\g \equiv g'_{0}(x=1) \sim u$ is the derivative of the starting
RSB function $g_{0}(x)$ at $x=1$.

Like in the RS case the renormalized parameters are asymptotically
approaching the pure system fixed point $\tl{g} = \e/12$, $g(x)=0$.
Nevertheless, the structure of the asymptotic solution for the
renormalized function $g(x;\xi)$ near this fixed point exhibits strong RSB.

However, the specific heat appears to be not effected by the RSB.
According to eq.(\ref{eef}) the leading singularity of the specific heat
is defined by the integral $\int_{0}^{1}dx g(x;\xi) \equiv \ol{g}(\xi)$
and not the function $g(x;\xi)$ itself.
It can be shown (see eq.(B.12)) that in the asymptotic regime
the value of $\ol{g}(\xi)$ coinsides with the RS asymptotics (\ref{zzzd}):

\be
\lb{zzzh}
\ol{g}(\xi) \sim \4 \xi^{-1}
\ee
Therefore, for the specific heat singularity one obtains the result
coinsiding with the RS one, eq.(\ref{zzzf}).

On the other hand, the asymptotic behaviour of the correlation functions,
appears to be quite different from the results of the traditional RS solution.
In the RS case, eq.(\ref{zzzd}), according to eq.(\ref{ddhh})
for the correlation function

\be
\lb{zzzii}
K(R) = \ol{\la\la\f(0)\f(R)\ra\ra^{2}}
\ee
one easily finds the following result:

\be
\lb{zzzi}
K(R) \sim (G_{0}(R))^{2}
\exp\{ 4 \int_{0}^{\ln R} d\xi g(\xi) \}  =
(G_{0}(R))^{2} \ln R
\ee
Therefore, in the RS case the disorder provides only the logarithmic
correction to the correlation function.

In the case of the RSB solution, eq.(\ref{bbbz}), according to
eq.(\ref{ddhh}) for the replica correlation function $K(x;R)$
one easily finds:

\be
\lb{zzzj}
K(x;R) \sim \left\{ \begin{array}{ll}
(G_{0}(R))^{2}\exp\{(const)\sqrt{\g\ln R}\} ; \; \; \; \;
(1-x)\sqrt{\g\ln R} >> 1  \\
(G_{0}(R))^{2} ; \; \; \; \; \; \; \; \; \; \; \; \; \; \; \; \; \; \; \; \; \;
\; \; \; \; \; \; \; \; \; \; \; \; \; \; \; (1-x)\sqrt{\g\ln R} << 1 \\
         \end{array}
         \right.
\ee
Correspondingly, for the "observable" correlation function,
eq.(\ref{zzzii}), one eventually obtains:

\be
\lb{zzzk}
K(R) = \int_{0}^{1}dx K(x;R) \sim
(G_{0}(R))^{2}\exp\{(const)\sqrt{\g\ln R}\}
\ee
This result is essentially different from the RS one, eq.(\ref{zzzi}).

\sectio{Discussion}

In this section we summarize our conclusions concerning the random
bond p-component Heisenberg ferromagnet and discuss the
remaining issues.

\vspace{5mm}

Spontaneous replica symmetry breaking coming from the interaction
of the fluctuations with the multiple local minima solutions of
the mean-field equations has a dramatic effect on
the renormalization group flows and on the critical
properties. In the systems with the number of spin components $p < 4$
the traditional RG flows at dimensions $D=4-\e$, which are usually
considered as describing the disorder-induced universal critical behavior,
appear to be unstable with respect to the RSB potentials as found
in spin glasses. For a general type of the Parisi RSB structures there
exists no stable fixed points, and the RG flows lead to the
{\it strong coupling regime} at the finite scale $R_{*} \sim \exp(1/u)$,
where $u$ is the small parameter describing the disorder.
Unlike the systems with $1 < p < 4$, where there exist stable
fixed points having 1-step RSB structures, eq.(\ref{bbbr}),
in the Ising case, $p=1$, there exist no stable fixed points,
and any RSB interactions lead to the strong coupling regime.

\vspace{5mm}

If there is RSB in the fourth-order potential, one
could identify a phase with a different symmetry than
the conventional paramagnetic phase, and thus there
would have to be a temperature $T_{RSB}$ at which
this change in symmetry occurs.
Actually, the RSB situation is the property of the
statistics of the saddle-point solutions only, and it is clear that
for large enough $\tau$ there
must be no RSB. Therefore, one can try to solve the
problem of summing over saddle-poin solutions for arbitrary $\tau$,
aiming to find finite value of $\tau_{c}$ at which the RSB solution
for this problem disappears.

Of course, in general this problem is very difficult to solve, but one
can easily obtain an estimate for the value of $\tau_{c}$
(assuming that at $\tau = 0$ the RSB situation takes place).
According to the qualitative study of this problem in the paper \cite{dhss},
the RSB solution
can occur only when the effective interactions between the "islands",
(where the system is effectively below $T_{c}$) are getting
non-small. The islands are the regions where $\delta\tau(r) > \tau$.
According to the Gaussian distribution for $\delta\tau(r)$,
the average distance between them must be
of the order of $\exp[-\t^{2}/u]$, so that the islands are
getting distant at $\tau > \sqrt{u}$. The interaction between the
islands is exponentially small in their separation. Therefore
at $\tau > \sqrt{u}$ they must be getting weakly interacting, and
there must be no RSB.

Note now that the shift of $T_{c}$ with respect to the corresponding
pure system is also of the order of $\sqrt{u}$. On the other hand,
the existence of local solutions to the mean-field equations remines
the Griffith phase \cite{griff} which is claimed to be observed in the
temperature interval between $T_{c}$ of the disordered system and $T_{c}$
of the corresponding pure system. On these grounds it is tempting to
associate the (hypothetical) RSB transition in the statistics
of the saddle-point solutions with the Griffith transition.
Correspondingly, it would also be natural to suggest that discovered
RSB phenomena in the scaling properties of weakly disordered systems
could be associated with the Griffith effects

\vspace{5mm}

The other key question which remains unanswered, is whether or
not the obtained strong coupling phenomena in the RG flows
could be interpreted as the onset of a kind of the spin-glass phase
near $\T$. Since it is the RSB interaction parameter describing disorder,
$g(x;\xi)$, which is the most divergent,
it is tempting to argue that in the temperature interval
$\t << \t_{*} \sim \exp(-1/u)$ near $\T$
the properties of the system should be essentially SG-like.

It should be stressed, however, that in the present study we observe only the
{\it crossover} temperature $\t_{*}$, at which the change of the
critical regime occurs, and it is hardly possible to associate this
temperature with any kind of phase transition. Therefore, if the RSB
effects could indeed provide any kind of true thermodynamic order
parameter, then this must be true in a whole temperature
interval where the RSB potentials exist.

The true spin-glass order (in the traditional sense) arises
from the onset of non-zero order parameter
$Q_{ab}(x) = <\phi_{a}(x)\phi_{b}(x)>; a\not= b$, and, at least
for the infinite-range model, $Q_{ab}$ develops the hierarchical
dependence on replica indices obtained by Parisi \cite{GP}.
In the present problem we only find that the coupling matrix $g_{ab}$
for the fluctuating fields develops strong RSB structure and its
elements are getting non-small at the finite scale.
Therefore, it seems more realistic to interprete discovered
RSB strong coupling phenomena in the RG just as a new type of the
critical behaviour characterized by strong SG-effects in the
scaling properties rather then in the ground state.

In spin-glasses it is generally believed that RSB phenomenon can be
interpreted as a factorization of the phase space into (ultrametric)
hierarchy of "valleys", or local minima pure states, separated by
macroscopic (infinite) barries \cite{sg}. Although in the systems
considered here the local minima configurations responcible for the RSB
are not likely to be separated by infinite barriers (otherwise it would mean
true SG freezing), it would be natural to interpret obtained phenomenon
as effective factorization of the phase space into a hierarchy
of valleys separated by {\it finite} barriers. Since the only
relevant scale in the critical region is the correlation length
the maximum energy barriers must be proportional to $\R^{D}(\t)$,
and they are getting divergent as the critical temperature is approached.
In this situation one could expect that besides the usual critical slowing
down (corresponding to the relaxation inside one valley) qualitatively much
bigger (exponentially large) relaxation times would be required
for overcoming barriers separating different valleys. Therefore,
the traditional measuremets (made at finite equilibration times)
can actually correspond to the equlibration within one valley only,
and not to the true thermal equilibrium. Then in a close vicinity of
the critical point different measuremets of the critical properties
of e.g. spatial correlation functions (in the same sample) would exhibit
different results as if the state of the system is getting effectively
"trapped" in different valleys, and thus the traditional spin-glass situation
will be observed.

\vspace{10mm}

{\large \bf Acknowledgments}

VD acknowledges Laboratiore de Physique Theorique
de Ecole Normale Superieure for hospitality.
Numerous fruitfull and encouraging discussions with M.Mezard and
Vl.S.Dotsenko are also grately acknowledged.

This work has been supported in part by the INTAS Grant No.1010-CT93-0027
and by the Grant of the Russian Fund for Fundamental Research 93-02-2081

\vspace{15mm}

\def\nonumsection#1
   {\vskip 24pt
    \noindent {\bf #1\par}
    \vskip 12pt
    \nobreak
    \noindent}

\nonumsection{Appendix A: The asymptotic solution for $p<4$ case}

In this Appendix we derive the asymptotic solution of the RG
eqs.(\ref{aabb}),(\ref{aacc}):

$$\lb{aab}
\fr{d}{d\xi}g(x) = (\e - (4+2p)\tl{g})g(x) + 4g^{2}(x)
-2pg(x)\int_{0}^{1}dy g(y) - p\int_{0}^{x} dy (g(x)-g(y))^{2}
\eqno(A1) $$

$$\lb {aac}
\fr{d}{d\xi}\tl{g} = \e\tl{g} - (8+p)\tl{g}^{2} + p\ol{g^{2}}
\eqno(A2) $$
(where $\ol{g^{2}} \equiv \int_{0}^{1}dx g^{2}(x)$) for the number of
components $p<4$.

It can be shown aposteriori that the term $(\e - (4+2p)\tl{g})g(x)$
in the eq.(A1) is irrelevant in the asymptotic regime.
So, consider the equation:

$$\lb{aad}
\fr{d}{d\xi}g(x) = 4g^{2}(x)
- 2pg(x)\int_{0}^{1}dy g(y) - p\int_{0}^{x} dy (g(x)-g(y))^{2}
\eqno(A3) $$

After taking derivative over $x$ and after simple transformations one gets:

$$\lb{aae}
\fr{d}{d\xi}g'(x) = 2p g'(x) [(\lm - 1)g(x) - \int_{x}^{1} dy (1-y)g'(y)]
\eqno(A4) $$
where $\lm = 4/p > 1$. Let us introduce:

$$\lb{aaf}
V(x) \equiv \int_{x}^{1}dy (1-y)g'(y)
\eqno(A5) $$
According to this definition one has:
$$\lb{aaff}
\ba
g'(x) = -\fr{1}{1-x} V'(x) \\
\\
g(x) = \int_{0}^{x}dy g'(y) = -\int_{0}^{x}dy\fr{1}{1-y} V'(y)
\ea
\eqno(A6) $$
Here for simplicity we consider the case $g(x=0) = 0$ (the behaviour of
the solution for $g(x=0) \not= 0$ in the asymptotic regime can be shown
to be qualitatively the same).

Then, for the eq.(\ref{aae}) after simple transformations we get:

$$\lb{aah}
\fr{d}{d\xi}V'(x) = -2p V'(x) [\int_{0}^{x}dy\fr{\lm-y}{1-y}V'(y) +\ol{g}(\xi)]
\eqno(A7) $$
where $\ol{g}(\xi) \equiv \int_{0}^{1}dx g(x,\xi) = \int_{0}^{1}dx(1-x)g'(x)
= V(x=0;\xi)$.

Let us define now:

$$\lb{aai}
W(x;\xi) = \int_{0}^{x} dy\fr{\lm-y}{1-y}V'(y)\\
\eqno(A8) $$
or
$$\lb{aaii}
V'(x) = \fr{1-x}{\lm-x} W'(x)
\eqno(A9) $$
{}From eq.(A7) one gets:

$$\lb{aaj}
\fr{d}{d\xi}W'(x) = -2p W'(x) [W(x) + \ol{g}(\xi)]
\eqno(A10) $$
Integrating over $x$ yields:

$$\lb{aak}
\fr{d}{d\xi}W(x) = -p W^{2}(x) -2p W(x) \ol{g}(\xi)
\eqno(A11) $$
(Here the integration constant is zero because $W(x=0) \equiv 0$).
This equation can be easily solved for any given function $\ol{g}(\xi)$:

$$\lb{aal}
W(x;\xi) = \fr{W_{0}(x) \exp[-2p\int_{0}^{\xi} d\eta\ol{g}(\eta)]}{
1 + p W_{0}(x) \int_{0}^{\xi} dt \exp[-2p\int_{0}^{t} d\eta\ol{g}(\eta)]}
\eqno(A12) $$
where:
$$\lb{aam}
W_{0}(x) \equiv W(x;\xi =0) = -\int_{0}^{x}dy (\lm - y)g_{0}'(y)
\eqno(A13) $$
and $g_{0}(x) \equiv g(x;\xi=0)$. Coming back through the definitions
(A8) and (A5) for the function $g(x;\xi)$ one gets:

$$\lb{aan}
g(x;\xi) = \int_{0}^{x}dy\fr{g_{0}'(y) \Theta(\xi)}{
[1 - p \int_{0}^{\xi} d\eta \Theta(\eta) \int_{0}^{y}dz(\lm-z)g_{0}'(z)]^{2}}
\eqno(A14) $$
where:
$$\lb{aao}
\Theta(\xi) = \exp[-2p\int_{0}^{\xi}d\eta \ol{g}(\eta)]
\eqno(A15) $$
Integrating $\int_{0}^{1}dx g(x;\xi) \equiv \ol{g}(\xi)$ one gets
the equation for the unknown function $\ol{g}(\xi)$:

$$\lb{aap}
\ol{g}(\xi) = \int_{0}^{1}dy\fr{(1-y)g_{0}'(y) \Theta(\xi)}{
[1 - p \int_{0}^{\xi} d\eta \Theta(\eta) \int_{0}^{y}dz(\lm-z)g_{0}'(z)]^{2}}
\eqno(A16) $$
Now the problem is to find the asymptotic behavior of $\ol{g}(\xi)$.

Let us introduce:

$$\lb{aaq}
G(\xi) \equiv \int_{0}^{\xi}d\eta \ol{g}(\eta)
\eqno(A17) $$
Integrating (A16) we obtain:

$$\lb{aar}
G(\xi) = \int_{0}^{1}dy\fr{(1-y)g_{0}'(y) A(\xi)}{
[1 - p A(\xi) \int_{0}^{y}dz(\lm-z)g_{0}'(z)]}
\eqno(A18) $$
where:

$$\lb{aas}
A(\xi) = \int_{0}^{\xi} d\eta \exp[-2pG(\eta)]
\eqno(A19) $$
Let us redefine

$$\lb{aass}
\psi(\xi) \equiv (A(\xi))^{-1} =
\frac{1}{\int_{0}^{\xi} d\eta \exp[-2pG(\eta)]}
\eqno(A20) $$

Then:

$$\lb{aat}
G(\xi) = \int_{0}^{1}dy\fr{(1-y)g_{0}'(y)}{
[\psi(\xi) - p \int_{0}^{y}dz(\lm-z)g_{0}'(z)]}
\eqno(A21) $$
Now, let us redefine again:

$$\lb{aau}
\psi(\xi) = p\int_{0}^{1}dy(\lm-y)g_{0}'(y) + \f(\xi)
\eqno(A22) $$
{}From eq.(A21) we get:

$$\lb{aav}
G(\xi) = \int_{0}^{1}dy\fr{(1-y)g_{0}'(y)}{
[p \int_{y}^{1}dz(\lm-z)g_{0}'(z) + \f(\xi)]}
\eqno(A23) $$
Assuming that $\f(\xi)$ is small, (A23) can be estimated as
follows:

$$\lb{aaw}
G(\xi) = G_{c} + \int_{0}^{1}dy(1-y)g_{0}'(y)[
\fr{1}{p \int_{y}^{1}dz(\lm-z)g_{0}'(z) + \f(\xi)} -
\fr{1}{p \int_{y}^{1}dz(\lm-z)g_{0}'(z)} ]
\eqno(A24) $$
or
$$\lb{aaww}
G(\xi) = G_{c} -\f(\xi) \int_{0}^{1}dy\fr{(1-y)g_{0}'(y)}{
[p \int_{y}^{1}dz(\lm-z)g_{0}'(z) + \f(\xi)][p \int_{y}^{1}dz(\lm-z)g_{0}'(z)]}
\eqno(A25) $$
where

$$\lb{aax}
G_{c} \equiv \int_{0}^{1}dy\fr{(1-y)g_{0}'(y)}{p
\int_{y}^{1}dz(\lm-z)g_{0}'(z)}
\eqno(A26) $$

For $\f(\xi) <<1$ the leading contribution in the integral in (A25)
comes from the vicinity of $y=1$. Assuming that $g_{0}'(y=1) = \gamma \not= 0$,
this contribution can be estimated as follows:

$$\lb{aay}
\ba
G(\xi) \simeq G_{c} -\f(\xi) \int_{...}^{1}dy\fr{(1-y)\gamma}{
[p\gamma (\lm-1)(1-y) + \f(\xi)] p\gamma (\lm-1)(1-y)} \simeq \\
\\
\simeq G_{c} -\fr{\f(\xi)}{\gamma p^{2}(\lm -1)^{2}}\ln\fr{1}{\f(\xi)}
\ea
\eqno(A27) $$
Such that, as $\f \to 0$, the value of $G(\xi)$ goes to finite value
$G_{c}$, but near this point the behavior of this function is
non-analytic.

Now let us assume that there exists sertain scale $\xi_{c}$,
such that $\f(\xi \to \xi_{c}) \to 0$, and consider the behavior
near $\xi_{c}$.
Coming back to the definition (A20) we can estimate:

$$\lb{aaz}
\ba
\psi(\xi) = [\int_{0}^{\xi}d\eta \exp(-2pG(\eta))]^{-1} = \\
\\
= (\int_{0}^{\xi_{c}}d\eta \exp(-2pG(\eta)) -
\int_{\xi}^{\xi_{c}}d\eta \exp(-2pG(\eta)))^{-1}  \simeq \\
\\
\simeq (\int_{0}^{\xi_{c}}d\eta \exp(-2pG(\eta)) -
\exp(-2pG_{c}) (\xi_{c} - \xi))^{-1} \simeq \\
\\
\simeq \fr{1}{\int_{0}^{\xi_{c}}d\eta \exp(-2pG(\eta))} +
\fr{\exp(-2pG_{c})}{[\int_{0}^{\xi_{c}}d\eta \exp(-2pG(\eta))]^{2}}
(\xi_{c} - \xi)
\ea
\eqno(A28) $$
Comparing this result with (A22), we find that:

$$\lb{aba}
\f(\xi) \simeq a(\xi_{c} - \xi)
\eqno(A29) $$
where the parameters $\xi_{c}$ and $a$ are defined by:

$$\lb{abb}
\fr{1}{\int_{0}^{\xi_{c}}d\eta \exp(-2pG(\eta))} =
p\int_{0}^{1}dy(\lm-y)g_{0}'(y)
\eqno(A30) $$
and

$$\lb{abc}
a = \fr{\exp(-2pG_{c})}{[\int_{0}^{\xi_{c}}d\eta \exp(-2pG(\eta))]^{2}} =
[p\int_{0}^{1}dy(\lm-y)g_{0}'(y)]^{2} \exp(-2pG_{c})
\eqno(A31) $$

Let us estimate the parameters $\xi_{c}$ and $a$ by the order of magnitude.
The characteristic value of the initial function
$g_{0}(x)$ is of the order of $u << 1$, which is the characteristic
value of the quenched disorder. If the initial function $g_{0}(x)$ does
not have special anomaly near $x=1$, then its derivative $\gamma$ must
also be of the order of $u$.
Then, the above integrals can be estimated as follows:

$$\lb{abd}
G_{c} = \int_{0}^{1}dy\fr{(1-y)g_{0}'(y)}{p \int_{y}^{1}dz(\lm-z)g_{0}'(z)}
\sim 1
\eqno(A32) $$
$$\lb{abe}
\int_{0}^{1}dy(\lm-y)g_{0}'(y) \sim u
\eqno(A33) $$
$$\lb{abf}
\int_{0}^{\xi_{c}}d\eta \exp(-2pG(\eta)) \sim \xi_{c}
\eqno(A34) $$

Thus, from (A30) and (A31) for the parameters $\xi_{c}$ and $a$
we find:

$$\lb{abg}
\xi_{c} \sim \fr{1}{u}
\eqno(A35) $$
$$\lb{abh}
a \sim u^{2}
\eqno(A36) $$

Now we can describe the qualitative behavior of the asymptotic solution.
According to (A27):

$$\lb{abi}
\ba
\ol{g}(\xi) = \fr{d}{d\xi} G(\xi) \simeq
\fr{a}{\gamma p^{2}(\lm -1)^{2}}\ln\fr{1}{(\xi_{c} - \xi)} \sim \\
\\
\sim u \ln\fr{1}{1 - u\xi}
\ea
\eqno(A37) $$
Therefore the value of the intergal $\int_{0}^{1}dx g(x;\xi) \equiv
\ol{g}(\xi)$
is formally getting divergent at finite scale $\xi_{c} \sim 1/u$.

Coming back to the result (A14) for the function $g(x;\xi)$ we have:

$$\lb{abj}
\ba
g(x;\xi) = \fr{ \Theta(\xi)}{(\int_{0}^{\xi} d\eta \Theta(\eta))^{2}}
\int_{0}^{x}dy\fr{g_{0}'(y)}{
[p\int_{y}^{1}dz(\lm-z)g_{0}'(z) + \f(\xi)]^{2}}\\
\\
= -[\fr{d}{d\xi}\fr{1}{\int_{0}^{\xi} d\eta \Theta(\eta)}]
\int_{0}^{x}dy\fr{g_{0}'(y)}{
[p\int_{y}^{1}dz(\lm-z)g_{0}'(z) + \f(\xi)]^{2}} \\
\\
= -[\fr{d}{d\xi} \psi(\xi)]
\int_{0}^{x}dy\fr{g_{0}'(y)}{
[p\int_{y}^{1}dz(\lm-z)g_{0}'(z) + \f(\xi)]^{2}} \\
\\
\simeq a \int_{0}^{x}dy\fr{g_{0}'(y)}{[p\int_{y}^{1}dz(\lm-z)g_{0}'(z) +
a(\xi_{c} - \xi)]^{2}}
\ea
\eqno(A38) $$
Therefore, when approaching the critical scale< $\xi \to \xi_{c}$,
the values of $g(x;\xi)$ are formally getting big in the narrow interval
$(1-x) << \D(\xi)$, where:

$$\lb{abkk}
\D(\xi) \sim \fr{a}{\gamma}(\xi_{c} - \xi) \sim (1 - u\xi)
\eqno(A39) $$
In this interval:

$$\lb{abk}
\ba
g(x;\xi) \simeq g(x=1;\xi) \equiv g_{1}(\xi) \simeq \\
\\
\simeq a \int_{...}^{1}dy\fr{\gamma}{[p(\lm-1)\gamma (1-y) +
a(\xi_{c} - \xi)]^{2}} \simeq \\
\\
\simeq \fr{1}{p(\lm -1)} \fr{1}{\xi_{c} - \xi} \\
\\
\sim a \fr{u}{1-u\xi}
\ea
\eqno(A40) $$
where $a \sim 1$ is a (non-universal) constant.

Therefore, the considered RG approach can be applied only up to the scales,
such that $(1-u\xi) \sim u$ (until the value of the parameter $g_{1}$
is getting non small).

Now let's come
back to the equation for the diagonal parameter $\tl{g}$ (A2).
According to the asymptotics obtained above we can estimate the value
of $\ol{g^{2}}$. Since the leading contibution comes from the region
$\D(\xi)$ near $x=1$, we get:

$$\lb{abl}
\ol{g^{2}} = \int_{0}^{1}dx g^{2}(x;\xi) \sim
(1-u\xi) \fr{a^{2}u^{2}}{(1-u\xi)^{2}} = a^{2}\fr{u^{2}}{1-u\xi}
\eqno(A41) $$
Therefore, from the equation (A2) we see that $\tl{g}$ diverges
as the logarithm:

$$\lb{abm}
\tl{g}(\xi) \sim u \ln\fr{1}{1-u\xi}
\eqno(A42) $$
Thus, in the region near $x=1$ where the value of $g(x;\xi)$
was obtained to be of the order of $u/(1-u\xi)$ the first term
$(\e - (4+2p)\tl{g})g(x)$ in the eq.(A1) is much smaller
than the other terms:

$$\lb{abn}
\tl{g}g(x) \sim \fr{u^{2}}{1-u\xi}\ln\fr{1}{1-u\xi} << \fr{u^{2}}{(1-u\xi)^{2}}
\sim g^{2}(x)
\eqno(A43) $$

Of course, the above asymptitic solution does not make possible
to obtain the behavior of the function $g(x;\xi)$ in the whole
interval $[0,1]$ for all $\xi$. Nevertheless, the numerical solution
of the general
RG equations (A1), (A2) clearly demonstrates that
at large scales the function $g(x;\xi)$ qiuckly goes to zero for all $x$
not too close to $1$, while in the narrow region near $x=1$ the values of this
function are getting divergent. Thus, the behaviour of the asymptotic solution
for $g(x;\xi)$ in the vicinity of the critical scale $\xi_{c}$
could be qualitatively represented as follows:

$$\lb{aannz}
g(x;\xi) \sim \left\{ \begin{array}{ll}
a\fr{u}{1 - u\xi} ; \; \; \; \mbox{for $(1-x) << \D(\xi)$}\\
\\
0 ; \; \; \;  \mbox{for $(1-x) >> \D(\xi)$}
\end{array}
\right.
\eqno(A44) $$
where $\D(\xi) = (1 - u\xi) \to u << 1$ as $\xi \to \xi_{c}$, and
$a$ is a positive non-universal constant.

\vspace{3mm}

The obtained asymptotics can also be
easily generalized for the situation when $g(x=0) \not= 0$.
One has to write: $g(x;\xi) = $ (the obtained solution) $ + g(x=0;\xi)$,
then put it into the equation, obtain the equation for $g(x=0;\xi) $,
and find the asymptotics for $g(x=0;\xi) $
It's straightforward to check that qualitatively
it dosn't change the above results.

\nonumsection{Appendix B: The asymptotic solution for $p=4$}

In $p=4$ case the asympotic solution of the equations (\ref{aabb})-(\ref{aacc})
can be obtained as follows.

Redefining the diagonal parameter $\tl{g}(\xi)$:

$$\lb{bbd}
\tl{g}(\xi) = \fr{\e}{12} + q(\xi)
\eqno(B1) $$
we get:

$$\lb{bbee}
\fr{d}{d\xi}g(x) = -12q(\xi)g(x) + 4g^{2}(x)
-8g(x)\int_{0}^{1}dy g(y) - 4\int_{0}^{x} dy (g(x)-g(y))^{2}
\eqno(B2) $$
$$\lb{bbe}
\fr{d}{d\xi} q(\xi) = -\e q - 12 q^{2} + 4 \ol{g^{2}}
\eqno(B3) $$

Then, proceeding like in the Appendix A, instead of eq.(A7)
we obtain:

$$\lb{bbaa}
\fr{d}{d\xi}V'(x;\xi) = -8 V'(x;\xi) V(x;\xi) - 12V'(x;\xi) q(\xi)
\eqno(B4) $$
Integration over $x$ yields:

$$\lb{bbab}
\fr{d}{d\xi}V(x;\xi) = -4 V^{2}(x;\xi) -12 q(\xi) V(x;\xi)
\eqno(B5) $$
(the integration constant is zero, since $V(x=1) \equiv 0$).
The solution of this equation for any given function $q(\xi)$ is:

$$\lb{bbac}
V(x;\xi) = \fr{V_{0}(x)\exp\{-12\int_{0}^{\xi}d\eta q(\eta)\}}{
1 + 4V_{0}(x)\int_{0}^{\xi}d\eta\exp\{-12\int_{0}^{\eta}dt q(t)\}}
\eqno(B6) $$
where $V_{0}(x)\equiv V(x;\xi=0) = \int_{x}^{1}dy(1-y)g_{0}'(x)$.

Coming back to the function $g(x;\xi)$ we get:

$$\lb{bbad}
g(x;\xi) = \int_{0}^{x}dy g'(y) + g(x=0;\xi)
= -\int_{0}^{x}dy \fr{1}{1-y}V'(y) + g(x=0;\xi)
\eqno(B7) $$
Using (B6) we find:

$$\lb{bba}
g(x;\xi) = \int_{0}^{x}dy \fr{g_{0}'(y)\exp\{-12\int_{0}^{\xi}d\eta q(\eta)\}}{
[1 + 4\int_{y}^{1}dz(1-z)g_{0}'(z)
\int_{0}^{\xi}d\eta\exp\{-12\int_{0}^{\eta}dt q(t)\}]^{2}}  + g(x=0;\xi)
\eqno(B8) $$
Putting this result back into the original equation (B2) we
get the equation for $g(x=0;\xi)$:

$$\lb{bbae}
\fr{d}{d\xi}g(x=0;\xi) =
 -12q(\xi)g(x=0;\xi) - 4g^{2}(x=0;\xi) - 8g(x=0;\xi)\ol{g}(\xi)
\eqno(B9) $$
where

$$\lb{bbaf}
\ol{g}(\xi) = \int_{0}^{1}dx\int_{0}^{x}dy
\fr{g_{0}'(y)\exp\{-12\int_{0}^{\xi}d\eta q(\eta)\}}{
[1 + 4\int_{y}^{1}dz(1-z)g_{0}'(z)
\int_{0}^{\xi}d\eta\exp\{-12\int_{0}^{\eta}dt q(t)\}]^{2}}
\eqno(B10) $$

Let us assume now that the parameter $q(\xi)$ decays as $\sim \xi^{-s}$
with $s > 1$. Then the integral $\int^{\xi} d\eta q(\eta)$ is converging
at large $\xi$, and for the exponent in (B8) we find that it is
equal to a constant of the order of one:
$\exp\{-12\int_{0}^{\xi}d\eta q(\eta)\} = A$.

Correspondingly, instead of eqs.(B8),(B10) we get:

$$\lb{bbaaz}
g(x;\xi) \simeq \int_{0}^{x}dy \fr{A g_{0}'(y)}{
[1 + 4A\xi\int_{y}^{1}dz(1-z)g_{0}'(z)]^{2}}  + g(x=0;\xi)
\eqno(B11) $$
and

$$\lb{bbafz}
\ba
\ol{g}(\xi) \simeq \int_{0}^{1}dx\int_{0}^{x}dy
\fr{A g_{0}'(y)}{
[1 + 4A\xi\int_{y}^{1}dz(1-z)g_{0}'(z)]^{2}} \\
\\
= \fr{A\ol{g}_{0}}{1 + 4A\xi\ol{g}_{0}}
\ea
\eqno(B12) $$
where
$\ol{g}_{0} \equiv \int_{0}^{1}dx g_{0}(x) = \int_{0}^{1}dx (1-x)g_{0}'(x)$.

Simple analycis of the integral in eq.(B11) shows that
actually it is the non-zero derivative $g_{0}'(x)$ near the point
$x=1$ which is important in the asymptotic regime. Whatever the function
$g_{0}(x)$ is in the region $(1-x) >> (\g\xi)^{-1/2}$,
it is always decaying like
$\xi^{-2}$ there, while for $(1-x) << (\g\xi)^{-1/2}$ the decay is
$(\g\xi)^{-1/2}$, where $\g = g_{0}'(x=1)$:

$$\lb{bbb}
g(x;\xi) \sim \left\{ \begin{array}{ll}
               \xi^{-2} ,  \; \; \; \; (1-x) >> \fr{1}{\sqrt{\g\xi}}  \\
               \fr{1}{\sqrt{\g\xi}} \; \; \; \; (1-x) << \fr{1}{\sqrt{\g\xi}}\\
                 \end{array}
                 \right.
\eqno(B13) $$
Besides, using (B12), from eq.(B9) one finds
that $g(x=0;\xi) \sim \xi^{-2}$.

Note now that according to the above asymptotic behaviour of the
function $g(x;\xi)$ at scales $\xi >> 1/u$ the leading contribution
to the quantity
$\ol{g^{2}} \equiv \int_{0}^{1} g^{2}(x;\xi)$ comes from the region
$(1-x) << \fr{1}{\sqrt{u\xi}}$:

$$\lb{bbc}
\ol{g^{2}} \sim \fr{1}{\sqrt{\xi}}(\fr{1}{\sqrt{\xi}})^{2} = \xi^{-3/2}
\eqno(B14) $$
Then, coming back to the eq.(B3) we find:

$$\lb{zzz}
q(\xi) \simeq \exp(-\e\xi)\int^{\xi} dt t^{-3/2} \exp(+\e t) \sim
a_{1}\xi^{-3/2} + a_{2}\xi^{-5/2} + ... \sim \xi^{-3/2}
\eqno(B15) $$
which is selfconsistent with
the assumption $q(\xi) \sim \xi^{-s}$ (with $s > 1$) made above.

Therefore, the asymptotic behaviour of the solution for $p=4$
at scales $\xi >> 1/u$ is given by eq.(B13).

\pagebreak


\begin{thebibliography}{99}

\bibitem{harris} A.B.Harris, J.Phys. {\bf C 7}, 1671 (1974)

\bibitem{new}
T. C. Lubensky and A. B. Harris, AIP Conf. Proc. {\bf 24} 311 (1974).
G. Grinstein, AIP Conf. Proc. {\bf 24} 313 (1974).
A. B. Harris and T. C. Lubensky, Phys.Rev.Lett., {\bf 33}, 1540 (1974);
G. Grinstein and A. Luther, Phys.Rev. {\bf B 13}, 1329 (1976).

\bibitem{newnew}
D. E. Khmelnitskii, ZhETF (Soviet Phys. JETP) {\bf 68}, 1960 (1975).

\bibitem{AA}
A. Aharony, in "Phase Transitions and Critical Phenomena,"
Ed.  C. Domb and J. Lebowitz, Vol. 6 (Academic Press, New York,
1976).

\bibitem{sg} M.Mezard, G.Parisi and M.Virasoro
"Spin-Glass Theory and Beyond", (World Scientific, Singapore, 1987).

\bibitem{manifold} M.Mezard and G.Parisi, J.Phys. I {\bf 1}, 809 (1991)

\bibitem{rsb-Marc} M.Mezard and A.P.Young, Europhys.Lett., {\bf 18}, 653 (1992)

\bibitem{Korshunov}
S. Korshunov, Phys.Rev. {\bf B48}, 3969 (1993)

\bibitem{dhss} Vik.S.Dotsenko, B.Harris, D.Sherrington and R.Stinchcombe,\\
Preprint cond-mat/9412106 (submitted to J.Phys.A: Math.Gen.)

\bibitem{rf} G.Parisi, Proceedings of the Les Houches 1982,
Session XXXIX, edited by J.B.Zuber and R.Stora (North Holland, Amsterdam)
1984

\bibitem{polyakov} A.M.Polyakov, "Gauge Fields and Strings",
Harwood Academic (1987)

\bibitem{griff} R.Griffiths, Phys.Rev.Lett., {\bf 23}, 17 (1969).
A.T.Ogielski, Phys.Rev., {\bf B32}, 7384 (1985).
A.J.Bray, Phys.Rev.Lett., {\bf 59}, 586 (1987)

\bibitem{rg}  G.Parisi "Statistical Field Theory", Addison-Wesley, 1988\\
A.Z.Patashinskii and V.L.Pokrovskii "Fluctuation Theory
of Phase Transitions", Pergamon Press, 1979

\bibitem{intro} Vik.S.Dotsenko "Introduction to the Theory of Spin-Glasses
and Neural Networks", World Scientific, 1994.

\bibitem{potts}
D.Gross, I.Kanter, and H.Sompolinsky, Phys.Rev.Lett. {\bf 55}, 304 (1985)

\bibitem{GP} G.Parisi, J.Phys. {\bf A13}, L115 (1980)





\end{thebibliography}
\end{document}